\documentclass[article]{jss}

\usepackage{bpk_jss_style}
\graphicspath{{figures/}}

\author{Brian P. Kent\\Carnegie Mellon University \And
		Alessandro Rinaldo\\Carnegie Mellon University \And
        Timothy Verstynen\\Carnegie Mellon University}
\title{\pkg{DeBaCl}: A \proglang{Python} Package for Interactive DEnsity-BAsed
CLustering}

\Plainauthor{Brian P. Kent, Alessandro Rinaldo, Timothy Verstynen}
\Plaintitle{DeBaCl: A Python Package for Interactive DEnsity-BAsed CLustering}

\Abstract{The level set tree approach of Hartigan~\citeyearpar{Hartigan1975}
provides a probabilistically based and highly interpretable encoding of the
clustering behavior of a dataset. By representing the hierarchy of data modes as
a dendrogram of the level sets of a density estimator, this approach offers many
advantages for exploratory analysis and clustering, especially for complex and
high-dimensional data. Several \proglang{R} packages exist for level set tree
estimation, but their practical usefulness is limited by computational
inefficiency, absence of interactive graphical capabilities and, from a
theoretical perspective, reliance on asymptotic approximations. To make it
easier for practitioners to capture the advantages of level set trees, we have
written the \proglang{Python} package \pkg{DeBaCl} for DEnsity-BAsed CLustering.
In this article we illustrate how \pkg{DeBaCl}'s level set tree estimates can be
used for difficult clustering tasks and interactive graphical data analysis. The
package is intended to promote the practical use of level set trees through
improvements in computational efficiency and a high degree of user
customization. In addition, the flexible algorithms implemented in \pkg{DeBaCl}
enjoy finite sample accuracy, as demonstrated in recent literature on density
clustering. Finally, we show the level set tree framework can be easily extended
to deal with functional data.} 

\Plainkeywords{density-based clustering, level set tree, Python, interactive
graphics, functional data analysis}

\Address{
  Brian P. Kent\\
  Department of Statistics\\ 
  Carnegie Mellon University\\
  Baker Hall 132\\
  Pittsburgh, PA 15213\\
  E-mail: \email{bpkent@andrew.cmu.edu}\\
  URL: \url{http://www.brianpkent.com}\\
  
  Alessandro Rinaldo\\
  Department of Statistics\\ 
  Carnegie Mellon University\\
  Baker Hall 132\\
  Pittsburgh, PA 15213\\
  E-mail: \email{arinaldo@cmu.edu}\\
  URL: \url{http://www.stat.cmu.edu/~arinaldo/}\\
  
  Timothy Verstynen\\
  Department of Psychology \& Center for the Neural Basis of Cognition\\
  Carnegie Mellon University\\
  Baker Hall 340U\\
  Pittsburgh, PA 15213\\
  E-mail: \email{timothyv@andrew.cmu.edu}\\
  URL: \url{http://www.psy.cmu.edu/~coaxlab/}
}

\begin{document}


\section{Introduction}\label{sec:intro}
Clustering is one of the most fundamental tasks in statistics and machine
learning, and numerous algorithms are available to practitioners. Some of the
most popular methods, such as K-means~\citep{MacQueen1967, Lloyd1982} and
spectral clustering~\citep{Shi2000}, rely on the key operational assumption that
there is one optimal partition of the data into $K$ well-separated groups, where
$K$ is assumed to be known {\it a priori}. While effective in some cases, this
flat or scale-free notion of clustering is inadequate when the data are very
noisy or corrupted, or exhibit complex multimodal behavior and spatial
heterogeneity, or simply when the value of $K$ is unknown. In these cases,
hierarchical clustering affords a more realistic and flexible framework in which
the data are assumed to have multi-scale clustering features that can be
captured by a hierarchy of nested subsets of the data. The expression of these
subsets and their order of inclusions---typically depicted as a
dendrogram---provide a great deal of information that goes beyond the original
clustering task. In particular, it frees the practitioner from the requirement
of knowing in advance the ``right'' number of clusters, provides a useful global
summary of the entire dataset, and allows the practitioner to identify and focus
on interesting sub-clusters at different levels of spatial resolution.

There are, of course, myriad algorithms just for hierarchical clustering.
However, in most cases their usage is advocated on the basis of
heuristic arguments or computational ease, rather than well-founded
theoretical guarantees. The high-density hierarchical clustering paradigm put
forth by Hartigan~\citeyearpar{Hartigan1975} is an exception. It is based on
the simple but powerful definition of clusters as the maximal connected
components of the super-level sets of the probability density specifying the
data-generating distribution. This formalization has numerous advantages: (1) it
provides a probabilistic notion of clustering that conforms to the intuition that
clusters are the regions with largest probability to volume ratio; (2) it
establishes a direct link between the clustering task and the fundamental
problem of nonparametric density estimation; (3) it allows for a clear
definition of clustering performance and consistency~\citep{Hartigan1981} that
is amenable to rigorous theoretical analysis and (4) as we show below, the
dendrogram it produces is highly interpretable, offers a compact yet informative
representation of a distribution, and can be interactively queried to extract
and visualize subsets of data at desired resolutions. Though the notion of
high-density clustering has been studied for quite some
time~\citep{Polonik1995}, recent theoretical advances have further demonstrated
the flexibility and power of density clustering. See, for example,
\citet{Rinaldo2012, Rinaldo2010, Kpotufe2011, Chaudhuri2010, Steinwart2011,
Sriperumbudur2012, Lei2013a, Balakrishnan2013} and the refences therein.

This paper introduces the \proglang{Python} package \pkg{DeBaCl} for efficient
and statistically-principled DEnsity-BAsed CLustering. \pkg{DeBaCl} is not the
first implementation of level set tree estimation and clustering; the
\proglang{R} packages \pkg{denpro}~\citep{Klemela2004},
\pkg{gslclust}~\citep{Stuetzle2010}, and \pkg{pdfCluster}~\citep{Azzalini2012}
also contain various level set tree estimators. However, they tend to be too
inefficient for most practical uses and rely on methods lacking rigorous
theoretical justification. The popular nonparametric density-based clustering
algorithm DBSCAN~\citep{Ester1996} is implemented in the \proglang{R} package
\pkg{fpc}~\citep{Hennig2013} and the \proglang{Python} library
\pkg{scikit-learn}~\citep{Pedregosa2011}, but this method does not provide an
estimate of the level set tree.

\pkg{DeBaCl} handles much larger datasets than existing software, improves
computational speed, and extends the utility of level set trees in three
important ways: (1) it provides several novel visualization tools to improve
the readability and interpetability of density cluster trees; (2) it offers a
high degree of user customization; and (3) it implements several recent
methodological advances. In particular, it enables construction of level set
trees for arbitrary functions over a dataset, building on the idea that level
set trees can be used even with data that lack a bona fide probability density
fuction. \pkg{DeBaCl} also includes the first practical implementation of the
recent, theoretically well-supported algorithm from \cite{Chaudhuri2010}.

\section{Level set trees}\label{sec:lsts}
Suppose we have a collection of points $\mathbb{X}_n = \{x_1,\ldots,x_n\}$ in $
\mbb{R}^d$, which we model as i.i.d. draws from an unknown probability
distribution with probability density function $f$ (with respect to Lebesgue
measure). Our goal is to identify and extract clusters of $\mathbb{X}_n$
without any \textit{a priori} knowledge about $f$ or the number of clusters.
Following the statistically-principled approach of \citet{Hartigan1975},
clusters can be identified as modes of $f$. For any threshold value $\lambda
\geq 0$, the $\lambda$-\textit{upper level set} of $f$ is
\begin{equation}\label{eq:L.lambda}
L_\lambda(f) = \{x \in \mbb{R}^d: f(x) \geq \lambda\}.
\end{equation}
The connected components of $L_\lambda(f)$ are called the $\lambda$-clusters of
$f$ and \textit{high-density clusters} are $\lambda$-clusters for any value of
$\lambda$. It is easy to see that $\lambda$-clusters associated with larger
values of $\lambda$ are regions where the ratio of probability content
to volume is higher. Also note that for a fixed value of $\lambda$, the
corresponding set of clusters will typically not give a partition of $\{x: f(x)
\geq 0\}$.   
 
The level set tree is simply the set of all high-density clusters. This
collection is a tree because it has the following property: for any
two high-density clusters $A$ and $B$, either $A$ is a subset of $B$, $B$ is a
subset of $A$, or they are disjoint. This property allows us to visualize the
level set tree with a dendrogram that shows all high-density clusters
simultaneously and can be queried quickly and directly to obtain specific
cluster assignments. Branching points of the dendrogram correspond to density
levels where two or more modes of the pdf, i.e. new clusters, emerge. Each
vertical line segment in the dendrogram represents the high-density clusters
within a single pdf mode; these clusters are all subsets of the cluster at the
level where the mode emerges. Line segments that do not branch are considered
high-density modes, which we call the leaves of the tree. For simplicity, we
tend to refer to the dendrogram as the level set tree itself.

Because $f$ is unknown, the level set tree must be estimated from the data.
Ideally we would use the high-density clusters of a suitable density estimate
$\wht{f}$ to do this; for a well-behaved $f$ and a large sample size, $\wht{f}$
is close to $f$ with high probability so the level set tree for $\wht{f}$ would
be a good estimate for the level set tree of $f$~\citep{Chaudhuri2010}.
Unfortunately, this approach is not computationally feasible even for
low-dimensional data because finding the upper level sets of $\wht{f}$ requires
evaluating the function on a dense mesh and identifying $\lambda$-clusters
requires a combinatorial search over all possible paths connecting any two
points in the mesh.

Many methods have been proposed to overcome these computational obstacles. The
first category includes techniques that remain faithful to the idea that
clusters are regions of the sample space. Members of this family include
histogram-based partitions~\citep{Klemela2004}, binary tree
partitions~\citep{Klemela2005} (implemented in the \proglang{R} package
\pkg{denpro}) and Delaunay triangulation partitions~\citep{Azzalini2007}
(implemented in \proglang{R} package \pkg{pdfCluster}). These techniques tend to
work well for low-dimension data, but suffer from the curse of dimensionality
because partitioning the sample space requires an exponentially increasing
number of cells or algorithmic complexity~\citep{Azzalini2007}.

In contrast, another family of estimators produces high-density clusters of data
points rather than sample space regions; this is the approach taken by our
package. Conceptually, these methods estimate the level set tree of $f$ by
intersecting the level sets of $f$ with the sample points $\mbb{X}_n$ and then
evaluating the connectivity of each set by graph theoretic means. This typically
consists of three high-level steps: estimation of the probability
density $\wht{f}(x)$ from the data; construction of a graph $G$ that describes
the similarity between each pair of data points; and a search for connected
components in a series of subgraphs of $G$ induced by removing nodes and/or
edges of insufficient weight, relative to various density levels.

The variations within the latter category are found in the definition of $G$,
the set of density levels over which to iterate, and the way in which $G$ is
restricted to a subgraph for a given density level $\lambda$. {\it Edge
iteration} methods assign a weight to the edges of $G$ based on the proximity of
the incident vertices in feature space~\citep{Chaudhuri2010} or the value of
$\wht{f}(x)$ at the incident vertices~\citep{Wong1983} or on a line segment
connecting them~\citep{Stuetzle2010}. For these procedures, the relevant density
levels are the edge weights of $G$. Frequently, iteration over these levels is
done by initializing $G$ with an empty edge set and adding successively more
heavily weighted edges, in the manner of traditional single linkage clustering.
In this family, the Chaudhuri and Dasgupta algorithm (which is a generalization
of Wishart~\citeyearpar{Wishart1969}) is particularly interesting because the
authors prove finite sample rates for convergence to the true level set
tree~\citep{Chaudhuri2010}. To the best of our knowledge, however,
only~\cite{Stuetzle2010} has a publicly available implementation, in the
\proglang{R} package \pkg{gslclust}.

{\it Point iteration} methods construct $G$ so the vertex for
observation $x_i$ is weighted according to $\wht{f}(x_i)$, but the edges are
unweighted. In the simplest form, there is an edge between the vertices for
observations $x_i$ and $x_j$ if the distance between $x_i$ and $x_j$
is smaller than some threshold value, or if $x_i$ and $x_j$ are among each
other's $k$-closest neighbors~\citep{Kpotufe2011, Maier2009}. A more complicated
version places an edge $(x_i, x_j)$ in $G$ if the amount of probability mass
that would be needed to fill the valleys along a line segment between $x_i$ and
$x_j$ is smaller than a user-specified threshold~\citep{Menardi2013}. The latter
method is available in the \proglang{R} package \pkg{pdfCluster}.

\section{Implementation}\label{sec:implement}
The default level set tree algorithm in \pkg{DeBaCl} is described in Algorithm
\ref{alg:debacl}, based on the method proposed by~\citet{Kpotufe2011}
and~\citet{Maier2009}. For a sample with $n$ observations in $\mbb{R}^d$, the
k-nearest neighbor (kNN) density estimate is:
\begin{equation}\label{eqn:knn_density}
	\wht{f}(x_j) = \frac{k}{n \cdot v_d \cdot r^d_k(x_j)}
\end{equation}
where $v_d$ is the volume of the Euclidean unit ball in $\mbb{R}^d$ and
$r_k(x_j)$ is the Euclidean distance from point $x_j$ to its $k$'th closest
neighbor. The process of computing subgraphs and finding connected components of
those subgraphs is implemented with the \pkg{igraph} package~\citep{Csardi2006}.
Our package also depends on the \pkg{NumPy} and \pkg{SciPy} packages for basic
computation~\citep{Jones2001} and the \pkg{Matplotlib} package for
plotting~\citep{Hunter2007}.
\begin{algorithm}
	\KwIn{$\{x_1,\ldots,x_n\}$, $k$, $\gamma$}
	\KwOut{$\wht{\mcr{T}}$, a hierarchy of subsets of $\{x_1,\ldots,x_n\}$}
	\BlankLine
	$G \leftarrow$ $k$-nearest neighbor similarity graph on
	$\{x_1,\ldots,x_n\}$\;
	$\wht{f}(\cdot) \leftarrow k$-nearest neighbor density estimate based on
	$\{x_1,\ldots,x_n\}$\;
	\For{$j \leftarrow 1$ \KwTo $n$}{
		$\lambda_j \leftarrow \wht{f}(x_j)$\;
		$L_{\lambda_j} \leftarrow \{x_i: \wht{f}(x_i) \geq \lambda_j\}$\;
		$G_j \leftarrow $ subgraph of $G$ induced by $L_j$\;
		Find the connected components of $G_{\lambda_j}$\;
	}
	$\wht{\mcr{T}} \leftarrow$ dendrogram of connected components of graphs
	$G_1,\ldots,G_n$, ordered by inclusions\;
	$\wht{\mcr{T}} \leftarrow$ remove components of size smaller than $\gamma$\;
	\Return{$\wht{\mcr{T}}$}
	\caption{Baseline \pkg{DeBaCl} level set tree estimation procedure}
	\label{alg:debacl}
\end{algorithm}
We use this algorithm because it is straightforward and fast; although it does
require computation of all $n \choose 2$ pairwise distances, the procedure can
be substantially shortened by estimating connected components on a sparse grid
of density levels. The implementation of this algorithm is novel in its own
right (to the best of our knowledge), and \pkg{DeBaCl} includes several other
new visualization and methodological tools.

\subsection{Visualization tools}\label{ssec:viz}
\begin{figure}[!ht]
	\centering
	\begin{subfigure}[b]{0.51\textwidth}
		\centering
		\includegraphics[width=\textwidth]{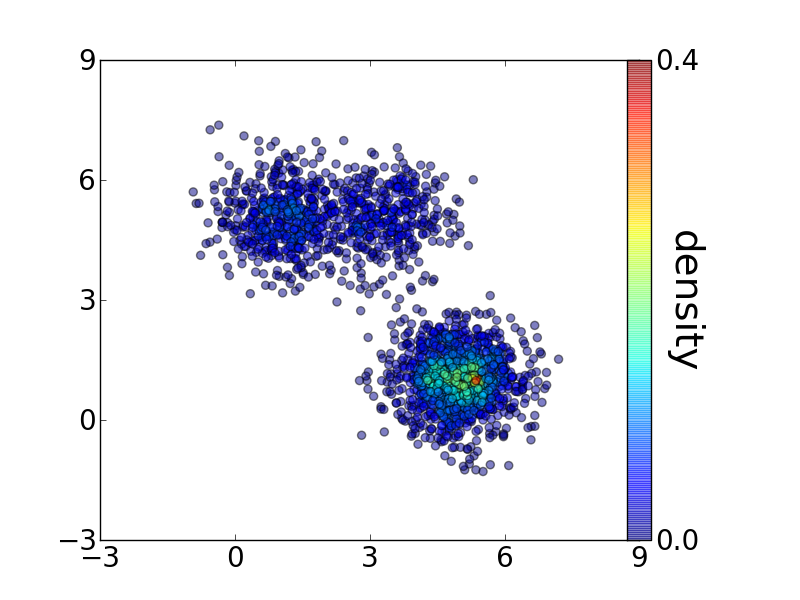}
		\caption{}
		\label{fig:gauss_fhat}
	\end{subfigure}
	\begin{subfigure}[b]{0.48\textwidth}
		\centering
		\includegraphics[width=\textwidth]{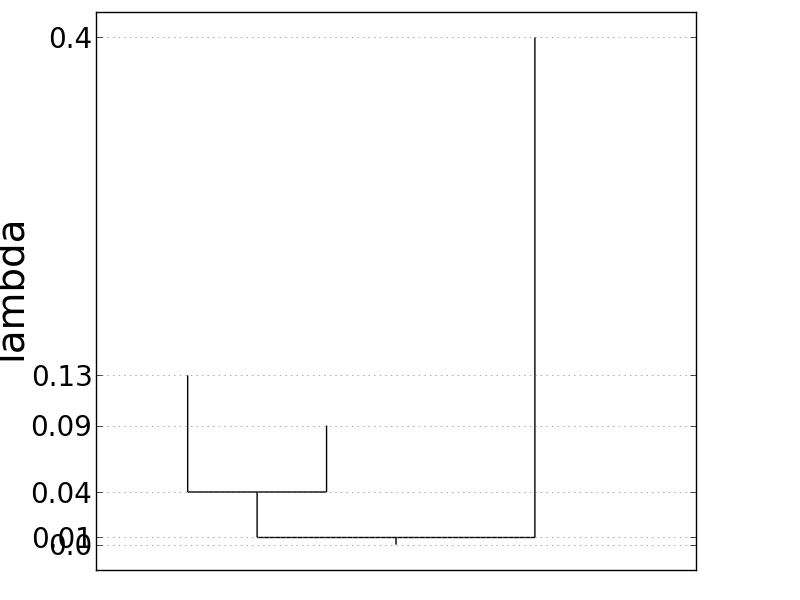}
		\caption{}
		\label{fig:gauss_tree_old}
	\end{subfigure}
	\begin{subfigure}[b]{0.49\textwidth}
		\centering
		\includegraphics[width=\textwidth]{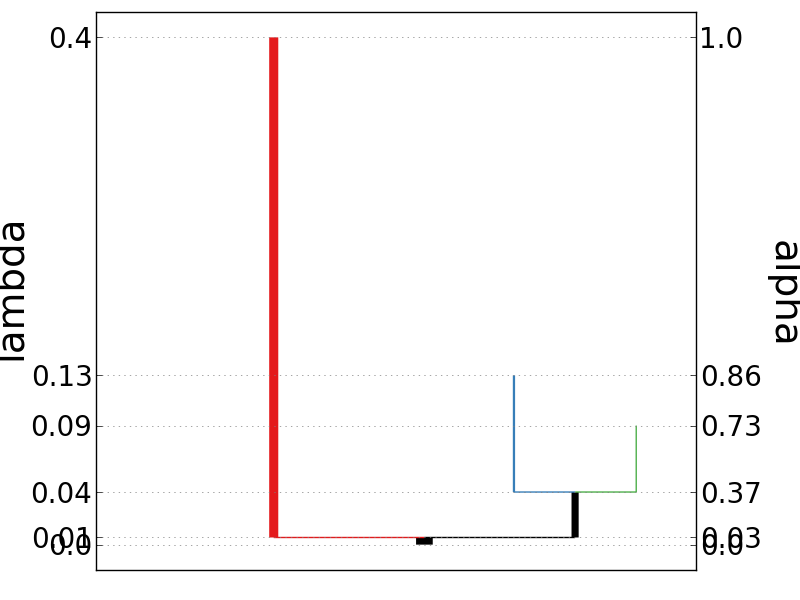}
		\caption{}
		\label{fig:gauss_tree_lambda}
	\end{subfigure}
	\begin{subfigure}[b]{0.49\textwidth}
		\centering
		\includegraphics[width=\textwidth]{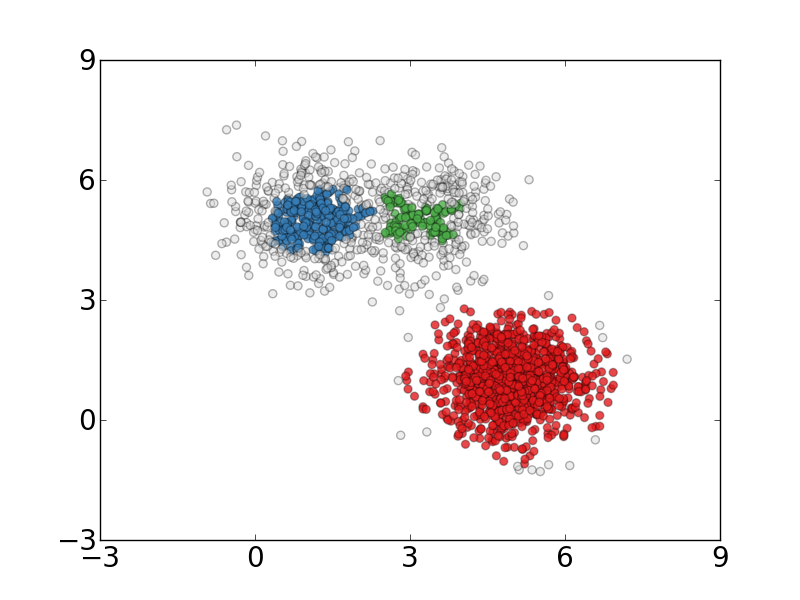}
		\caption{}
		\label{fig:gauss_allMode}
	\end{subfigure}
	\caption{Level set tree plots and cluster labeling for a simple simulation.
	A level set tree is constructed from a sample of 2,000 observations drawn
	from a mix of three Gaussians in $\mathbb{R}^2$. \subref{fig:gauss_fhat})
	The kNN density estimator evaluated on the data.
	\subref{fig:gauss_tree_old}) A plot of the tree based only on the
	mathematical definition of level set trees. \subref{fig:gauss_tree_lambda})
	The new level set tree plot, from \pkg{DeBaCl}. Tree branches emphasize
	empirical mass through ordering, spacing, and line width, and they are
	colored to match the cluster labels in \subref{fig:gauss_allMode}. A second
	vertical axis is added that indicates that fraction of background mass at
	each critical density level. \subref{fig:gauss_allMode}) Cluster labels from
	the \textit{all-mode} labeling technique, where each leaf of the level set
	tree is designated as a cluster.}
	\label{fig:gauss}
\end{figure}

Our level set tree plots increase the amount of information contained in a tree
visualization and greatly improve interpretability relative to existing
software. Suppose a sample of 2,000 observations in $\mbb{R}^2$ from a mixture
of three Gaussian distributions (Figure \ref{fig:gauss_fhat}). The traditional
level set tree is illustrated in Figure \ref{fig:gauss_tree_old} and the
\pkg{DeBaCl} version in Figure \ref{fig:gauss_tree_lambda}. A plot based only on
the mathematical definition of a level set tree conveys the structure of the
mode hierarchy and indicates the density levels where each tree node begins and
ends, but does not indicate how many points are in each branch or visually
associate the branches with a particular subset of data. In the proposed
software package, level set trees are plotted to emphasize the empirical mass in
each branch (i.e. the fraction of data in the associated cluster): tree
branches are sorted from left-to-right by decreasing empirical mass, branch
widths are proportional to empirical mass, and the white space around the
branches is proportional to empirical mass. For matching tree nodes to the data,
branches can be colored to correspond to high-density data clusters (Figures
\ref{fig:gauss_tree_lambda} and \ref{fig:gauss_allMode}). Clicking on a tree
branch produces a banner that indicates the start and end levels of the
associated high-density cluster as well as its empirical mass (Figure
\ref{fig:endpt_subtree}).

The level set tree plot is an excellent tool for interactive exploratory
data analysis because it acts as a handle for identifying and plotting
spatially coherent, high-density subsets of data. The full power of this feature
can be seen clearly with the more complex data of Section \ref{sec:usage}.

\subsection{Alternate scales}\label{ssec:scales}
By construction, the nodes of a level set tree are indexed by density levels
$\lambda$, which determine the scale of the vertical axis in a plot of the tree.
While this does encode the parent-child relationships in the tree,
interpretability of the $\lambda$ scale is limited by the fact that it depends
on the height of the density estimate $\wht{f}$. It is not clear, for example,
whether $\lambda = 1$ would be a low- or a high-density threshold; this depends
on the particular distribution.

To remove the scale dependence we can instead index level set tree nodes based
on the probability content of upper level sets. Specifically, let $\alpha$ be a
number between $0$ and $1$ and define 
\begin{equation}
	\lambda_\alpha = \sup \left\{ \lambda \colon \int_{x \in L_\lambda(f)} f(x)
	dx \geq \alpha \right\}
\end{equation}
to be the value of $\lambda$ for which the upper level set of $f$ has
probability content no smaller than $\alpha$~\citep{Rinaldo2012}. The map
$\alpha \mapsto \lambda_\alpha$ gives a monotonically decreasing one-to-one
correspondence between values of $\alpha$ in $[0,1]$ and values of $\lambda$ in
$[0,\max_x f(x)]$. In particular, $\lambda_1 = 0$ and $\lambda_0 = \max_x f(x)$.
For an empirical level set tree, set $\lambda_\alpha$ to the $\alpha$-quantile
of $\{\wht{f}(x_i)\}_{i=1}^n$. Expressing the height of the tree in terms of
$\alpha$ instead of $\lambda$ does not change the topology (i.e. number and
ordering of the branches) of the tree; the re-indexed tree is a deformation of
the original tree in which some of its nodes are stretched out and others are
compressed.

$\alpha$-indexing is more interpretable and useful for several reasons. The
$\alpha$ level of the tree indexes clusters corresponding to the $1 - \alpha$
fraction of ``most clusterable" data points; in particular, larger $\alpha$
values yield more compact and well-separated clusters, while smaller values can
be used for de-noising and outlier removal. Because $\alpha$ is always between
$0$ and $1$, scaling by probability content also enables comparisons of level
set trees arising from data sets drawn from different pdfs, possibly in spaces
of different dimensions. Finally, the $\alpha$-index is more effective than
$\lambda$-indexing in representing regions of large probability content but low
density and is less affected by small fluctuations in density estimates.

A common (incorrect) intuition when looking at an $\alpha$-indexed level set
tree plot is to interpret the height of the branches as the size of the
corresponding cluster, as measured by its empirical mass. However, with
$\alpha$-indexing the height of any branch depends on its empirical mass as
well as the empirical mass of all other branches that coexist with it. In order
to obtain trees that do conform to this intuition, we introduce the
$\kappa$-indexed level set tree.

Recall from Section \ref{sec:lsts} that clusters are defined as maximal
connected components of the sets $L_{\lambda}(f)$ (see equation
\ref{eq:L.lambda}) as $\lambda$ varies from $0$ to $\max_x f(x)$, and that the
level set tree is the dendrogram representing the hierarchy of all clusters.
Assume the tree is binary and with tooted. Let $\{1,2,\ldots, K\}$ be an
enumeration of the nodes of the level set tree and let $\mathcal{C} = \{
C_0,\ldots, C_K\}$ be the corresponding clusters. We can always choose the
enumeration in a way that is consistent with the hierarchy of inclusions of the
elements of $\mathcal{C}$; that is, $C_0$ is the support of $f$ (which we assume
for simplicity to be a connected set) and if $C_i \subset C_j$, then $i>j$. For
a node $i>0$, we denote with $\mathrm{parent}_i$ the unique node $j$ such that
$C_j$ is the smallest element of $\mathcal{C}$ such that $C_j \supset C_i$.
Similarly, $\mathrm{kid}_i$ is the pair of nodes $(j,j')$ such that $C_j$ and
$C_{j'}$ are the maximal subsets of $C_i$. Finally, for $i>0$, $\mathrm{sib}_i$
is the node $j$ such there exists a $k$ for which $\mathrm{kid}_k = (i,j)$. For
a cluster $C_i \in \mathcal{C}$, we set 
\begin{equation}
	M_i = \int_{C_i} f(x)dx,
\end{equation}
which we refer to as the {\it mass} of $C_i$.

The true $\kappa$-tree can be defined recursively by associating with each node
$i$ two numbers $\kappa'_i$ and $\kappa''_i$ such that $\kappa'_i - \kappa''_i$
is the {\it salient mass} of node $i$. For leaf nodes, the salient mass is the
mass of the cluster, and for non-leaves it is the mass of the cluster boundary
region. $\kappa'$ and $\kappa''$ are defined differently for each node type.
\begin{enumerate}
	\item Internal nodes, including the root node.
	\begin{gather*}
		\kappa'_0 = M_0 = 1,\\
		\kappa'_i = \kappa''_{\mathrm{parent}_i}\\
		\kappa''_i = \sum_{j \in \mathrm{kid}_i} M_j + \sum_{k \in
		\mathrm{sib}_i} M_k
	\end{gather*}
	
	\item Leaf nodes.
	\begin{gather*}
		\kappa'_i = \kappa''_{\mathrm{parent}_i}\\
		\kappa''_i = \kappa'_i - M_i
	\end{gather*}
\end{enumerate}

To estimate the $\kappa$-tree, we
use $\widehat{f}$ instead of $f$ and let
$m_i$ be the fraction of data contained in the cluster for the tree node $i$ at
birth. Again, define the estimated tree recursively:
\begin{gather*}
	\wht{\kappa}'_0 = 1,\\
	\wht{\kappa}'_i = \wht{\kappa}''_{\mathrm{parent}_i},\\
	\wht{\kappa}''_i = \wht{\kappa}'_i - m_i + \sum_{j \in
	\mathrm{kid_i}} m_j.
\end{gather*}
In practice we subtract the above quantities from 1 to get an increasing scale
that matches the $\lambda$ and $\kappa$ scales. 

\begin{figure}[!ht]
	\centering
	\begin{subfigure}[b]{0.4\textwidth}
		\centering
		\includegraphics[width=\textwidth]{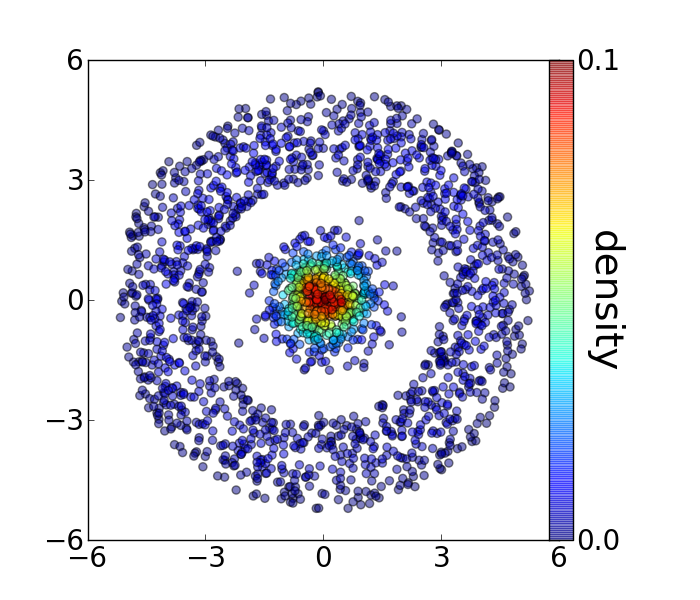}
		\caption{}
		\label{fig:crater_fhat}
	\end{subfigure}
	\begin{subfigure}[b]{0.345\textwidth}
		\centering
		\includegraphics[width=\textwidth]{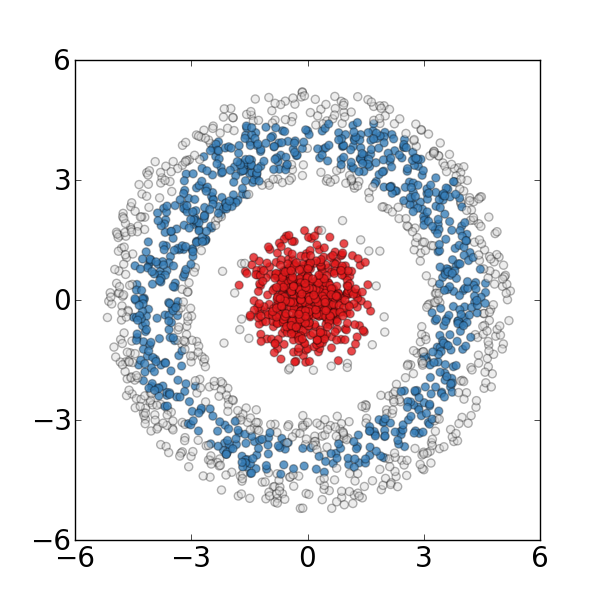}
		\caption{}
		\label{fig:crater_allMode}
	\end{subfigure}
	\caption{The crater simulation. 2,000 points are sampled from a mixture of a
	central Gaussian and an outer ring (Gaussian direction with uniform noise).
	Roughly 70\% of the points are in the outer ring. \subref{fig:crater_fhat})
	The kNN density estimator evaluated on the data.
	\subref{fig:crater_allMode}) Cluster labels from the \textit{all-mode}
	labeling technique, where each leaf of the level set tree is designated as a
	cluster. Gray points are unlabeled low-density background observations.}
	\label{fig:crater_data}
\end{figure}

\begin{figure}[!ht]
	\centering
	\begin{subfigure}[b]{0.325\textwidth}
		\centering
		\includegraphics[width=\textwidth]{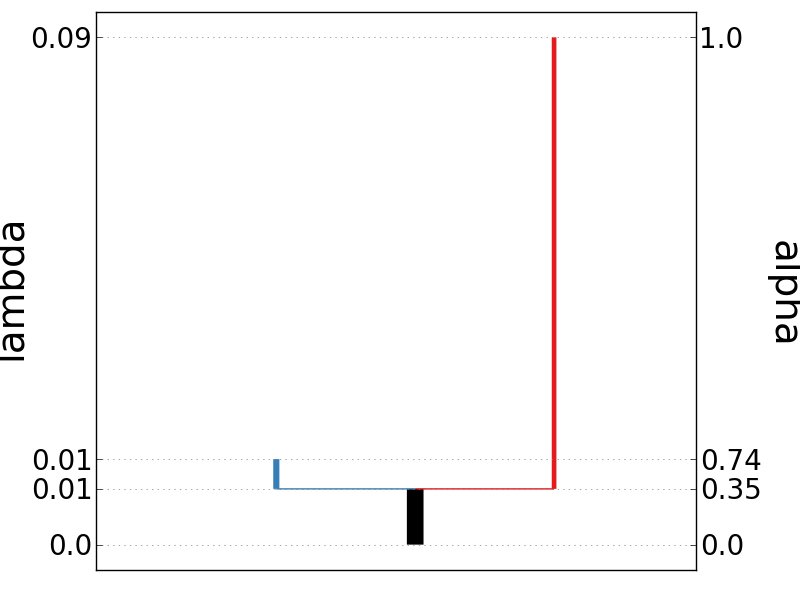}
		\caption{}
		\label{fig:crater_lambda}
	\end{subfigure}
	\begin{subfigure}[b]{0.325\textwidth}
		\centering
		\includegraphics[width=\textwidth]{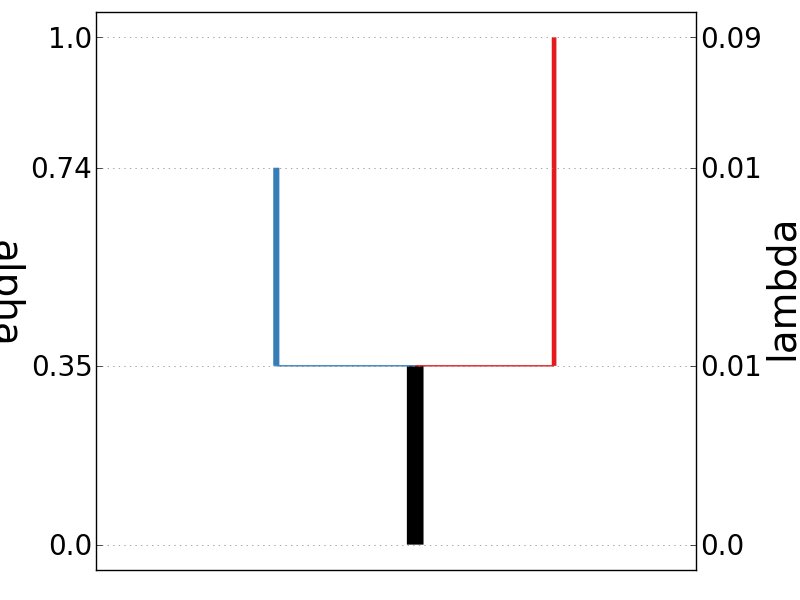}
		\caption{}
		\label{fig:crater_alpha}
	\end{subfigure}
	\begin{subfigure}[b]{0.325\textwidth}
		\centering
		\includegraphics[width=\textwidth]{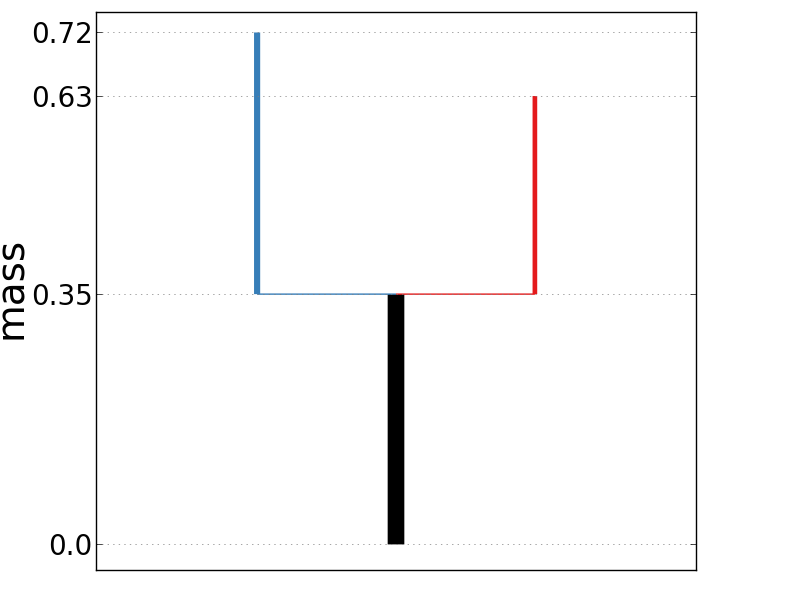}
		\caption{}
		\label{fig:crater_kappa}
	\end{subfigure}
	\caption{Level set tree scales for the crater simulation.
	\subref{fig:crater_lambda}) The $\lambda$ scale is dominant, corresponding
	directly to density level values. There is a one-to-one correspondence with
	$\alpha$ values shown on the right y-axis. Note the blue branch,
	corresponding to the outer ring in the crater simulation, appears to be very
	small in this plot, despite the fact that the true group contains about 70\%
	of data \subref{fig:crater_alpha}) The $\alpha$ scale is dominant,
	corresponding to the fraction of data excluded from the upper level set at
	each $\lambda$ value. The blue cluster is more exaggerated but the topology
	of the tree remains unchanged. \subref{fig:crater_kappa}) The $\kappa$
	scale. The blue cluster now appears larger than the red, facilitating the
	intuitive connection between branch height and cluster mass.}
	\label{fig:crater_trees}
\end{figure}

Note that switching between the $\lambda$ to $\alpha$ index does not change the
overall shape of the tree, but switching to the $\kappa$ index {\it does}. In
particular, the tallest leaf of the $\kappa$ tree corresponds to the cluster
with largest empirical mass. In both the $\lambda$ and $\alpha$ trees, on the
other hand, leaves correspond to clusters composed of points with high density
values. The difference can be substantial. Figure \ref{fig:crater_trees}
illustrates the differences between the three types of indexing for the
``crater'' example in Figure \ref{fig:crater_data}. This example consists of a
central Gaussian with high density and low mass surrounded by a ring with high
mass but uniformly low density. The $\lambda$-scale tree (Figure
\ref{fig:crater_lambda}) correctly indicates the heights of the modes of
$\wht{f}$, but tends to produce the incorrect intuition that the ring (blue node
and blue points in Figure \ref{fig:crater_allMode}) is small. The $\alpha$-scale
plot (Figure \ref{fig:crater_alpha}) ameliorates this problem by indexing node
heights to the quantiles of $\wht{f}$. The blue node appears at $\alpha = 0.35$,
when 65\% of the data remains in the upper level set, and vanishes at $\alpha =
0.74$, when only 26\% of the data remains in the upper level set. It is tempting
to say that this means the blue node contains $0.74 - 0.35 = 0.39$ of the mass
but this is incorrect because some of the difference in mass is due to the red
node. This interpretation is precisely the design of the $\kappa$-tree, however,
where we can say that the blue node contains $0.72 - 0.35 = 0.37$ of the data.

\subsection{Cluster retrieval options}\label{ssec:labeling}
Many clustering algorithms are designed to only output a partition of the data,
whose elements are then taken to be the clusters. As we argued in the
introduction, such a paradigm is often inadequate for data exhibiting complex
and multi-scale clustering features. In contrast, hierarchical clustering in
general and level set tree clustering in particular give a more complete and
informative description of the clusters in a dataset. However, many applications
require that each data point be assigned to a single cluster label. Much of the
work on level set trees ignores this phase of a clustering application or
assumes that labels will be assigned according to the connected components at a
chosen $\lambda$ (density) or $\alpha$ (mass) level, which \pkg{DeBaCl}
accomodates through the \textit{upper set clustering} option. Rather than
choosing a single density level, a practitioner might prefer to specify the
number of clusters $K$ (as with $K$-means). One way (of many) that this can be
done is to find the first $K-1$ splits in the level set tree and identify each
of the children from these splits as a cluster, known in \pkg{DeBaCl} as the
\textit{first-K clustering} technique. A third, preferred, option avoids the
choice of $\lambda$, $\alpha$, or $K$ altogether and treats each leaf of the
level set tree as a separate cluster~\citep{Azzalini2007}. We call this the
\textit{all-mode clustering} method. Use of these labeling options is
illustrated in Section \ref{sec:usage}.

Note that each of these methods assigns only a fraction of points to clusters
(the \textit{foreground} points), while leaving low-density observations
(\textit{background} points) unlabeled. Assigning the background points to
clusters can be done with any classification algorithm, and \pkg{DeBaCl}
includes a handful of simple options, including a k-nearest neighbor classifer,
for the task.

\subsection{Chaudhuri and Dasgupta algorithm}\label{ssec:cdTree}
Chaudhuri and Dasgupta~\citeyearpar{Chaudhuri2010} introduce an algorithm for
estimating a level set tree that is particularly notable because the authors
prove finite-sample convergence rates (where consistency is in the sense of~\citet{Hartigan1981}). The algorithm is a generalization of
single linkage, reproduced here for convenience in Algorithm \ref{alg:cd_tree}.
\begin{algorithm}
	\KwIn{$\{x_1,\ldots,x_n\}$, $k$, $\alpha$}
	\KwOut{$\wht{\mcr{T}}$, a hierarchy of subsets of $\{x_1,\ldots,x_n\}$}
	\BlankLine
	$r_k(x_i) \leftarrow$ distance to the $k$'th neighbor of $x_i$\; 
	\For{$r \leftarrow 0$ \KwTo $\infty$}{
		$G_r \leftarrow$ graph with vertices $\{x_i: r_k(x_i) \leq r\}$ and
		edges $\{(x_i, x_j): \| x_i - x_j \| \leq \alpha r\}$\;
		Find the connected components of $G_{\lambda_r}$\;
	}
	$\wht{\mcr{T}} \leftarrow$ dendrogram of connected components of graphs
	$G_r$, ordered by inclusions\;
	\Return{$\wht{\mcr{T}}$}
	\caption{\citet{Chaudhuri2010} level set tree estimation procedure.}
	\label{alg:cd_tree}
\end{algorithm}
To translate this program into a practical implementation, we must find a finite
set of values for $r$ such that the graph $G_r$ can only change at these values.
When $\alpha=1$, the only values of $r$ where the graph can change are the edge
lengths in the graph $e_{ij} = \|x_i - x_j\|$ for all $i$ and $j$. Let $r$ take
on each value of $e_{ij}$ in descending order; in each iteration remove vertices
and edges with larger k-neighbor radius and edge length, respectively.

When $\alpha \neq 1$, the situation is trickier. First, note that including $r$
values where the graph \textit{does not} change is not a problem, since the
original formulation of the method includes all values of $r \in \mbb{R}^{+0}$.
Clearly, the vertex set can still change at any edge length $e_{ij}$. The edge
set can only change at values where $r = e_{ij}/\alpha$ for some $i, j$. Suppose
$e_{u,v}$ and $e_{r,s}$ are consecutive values in a descending ordered list of
edge lengths. Let $r = e / \alpha$, where $e_{u,v} < e < e_{r,s}$. Then the edge
set $E = \{(x_i, x_j): \|x_i - x_j\| \leq \alpha r = e\}$ does not change as $r$
decreases until $r = e_{u,v}/\alpha$, where the threshold of $\alpha r$ now
excludes edge $(x_u, x_v)$. Thus, by letting $r$ iterate over the values in
$\bigcup_{i,j} \{e_{ij}, \frac{e_{ij}}{\alpha}\}$, we capture all possible
changes in $G_r$.

In practice, starting with a complete graph and removing one edge at a time is
extremely slow because this requires $2 * {n \choose 2}$ connected component
searches. The \pkg{DeBaCl} implementation includes an option to initialize the
algorithm at the k-nearest neighbor graph instead, which is a substantially
faster approximation to the Chaudhuri-Dasgupta method. This shortcut is still
dramatically slower than \pkg{DeBaCl}'s geometric tree algorithm, which is one
reason why we prefer the latter. Future development efforts will focus on
improvements in the speed of both procedures.

\subsection{Pseudo-densities for functional data}\label{ssec:pseudoDens} The
level set tree estimation procedure in Algorithm \ref{alg:debacl} can be
extended to work with data sampled from non-Euclidean spaces that do not admit a
well-defined pdf. The lack of a density function would seem to be an
insurmountable problem for a method defined on the levels of a pdf. In this
case, however, level set trees can be built on the levels of a
pseudo-density estimate that measures the similarity of observations and the
overall connectivity of the sample space. Pseudo-densities cannot be used to
compute probabilities as in Euclidean spaces, but are proportional to the
statistical expectations of estimates of the form $\wht{f}$, which remain
well-defined random quantities~\citep{Ferraty2006}.

Random functions, for example, may have well-defined probability distributions
that cannot be represented by pdfs~\citep{Billingsley2012}. To build level set
trees for this type of data, \pkg{DeBaCl} accepts very general functions for
$\wht{f}$, including pseudo-densities, although the user must compute the
pairwise distances. The package includes a utility function for evaluating a
k-nearest neighbor pseudo-density estimator on the data based on the pairwise
distances. Specifically, equation \ref{eqn:knn_density} is modified by expunging
the term $v^d$ and setting $d$ arbitrarily to 1. An application is shown in
Section \ref{sec:usage}.

\subsection{User customization}\label{ssec:customize}
One advantage of \pkg{DeBaCl} over existing cluster tree software is that
\pkg{DeBaCl} is intended to be easily modified by the user. As described above,
two major algorithm types are offered, as well as the ability to use
pseudo-densities for functional data. In addition, the package allows a high
degree of customization in the type of similarity graph, \textit{data ordering
function} (density, pseudo-density, or arbitrary function), pruning function,
cluster labeling scheme, and background point classifier. In effect, the only
fixed aspect of \pkg{DeBaCl} is that clusters are defined for every level to be
connected components of a geometric graph.

\section{Usage} \label{sec:usage}
\subsection{Basic Example}\label{ssec:example}

In this section we walk through the density-based clustering analysis of
10,000 fiber tracks mapped in a human brain with diffusion-weighted
imaging. For this analysis we use only the subcortical endpoint of each
fiber track, which is in $\mathbb{R}^3$. Despite this straightforward
context of finite, low-dimensional data, the clustering problem is
somewhat challenging because the data are known to have complicated
striatal patterns. For this paper we add the \pkg{DeBaCl} package to the
Python path at run time, but this can be done in a more persistent
manner for repeated use. The \pkg{NumPy} library is also needed for this
example, and we assume the dataset is located in the working directory.
We use our preferred algorithm, the geometric level set tree, which is
located in the \pkg{geom\_tree} module.

\begin{codecell}
\begin{codeinput}
\begin{lstlisting}
## Import DeBaCl package
import sys
sys.path.append('/home/brian/Projects/debacl/DeBaCl/')

from debacl import geom_tree as gtree
from debacl import utils as utl

## Import other Python libraries
import numpy as np

## Load the data
X = np.loadtxt('0187_endpoints.csv', delimiter=',')
n, p = X.shape
\end{lstlisting}
\end{codeinput}

\end{codecell}
The next step is to define parameters for construction and pruning of
the level set tree, as well as general plot aesthetics. For this example
we set the density and connectivity smoothness parameter $k$ to $0.01n$
and the pruning parameter $\gamma$ is set to $0.05n$. Tree branches with
fewer points than this will be merged into larger sibling branches. For
the sake of speed, we use a small subsample in this example.

\begin{codecell}
\begin{codeinput}
\begin{lstlisting}
## Downsample
n_samp = 5000
ix = np.random.choice(range(n), size=n_samp, replace=False)
X = X[ix, :]
n, p = X.shape

## Set level set tree parameters
p_k = 0.01
p_gamma = 0.05

k = int(p_k * n)
gamma = int(p_gamma * n)

## Set plotting parameters
utl.setPlotParams(axes_labelsize=28, xtick_labelsize=20, ytick_labelsize=20, figsize=(8,8))
\end{lstlisting}
\end{codeinput}

\end{codecell}
For straightforward cases like this one, we use a single convenience
function to do density estimation, similarity graph definition, level
set tree construction, and pruning. In the following example, each of
these steps will be done separately. Note the \texttt{print} function is
overloaded to show a summary of the tree.

\begin{codecell}
\begin{codeinput}
\begin{lstlisting}
## Build the level set tree with the all-in-one function
tree = gtree.geomTree(X, k, gamma, n_grid=None, verbose=False)
print tree
\end{lstlisting}
\end{codeinput}
\begin{codeoutput}

\begin{verbatim}
     alpha1  alpha2  children   lambda1   lambda2 parent  size
key                                                           
0    0.0000  0.0040    [1, 2]  0.000000  0.000003   None  5000
1    0.0040  0.0716  [11, 12]  0.000003  0.000133      0  2030
2    0.0040  0.1278  [21, 22]  0.000003  0.000425      0  2950
11   0.0716  0.3768  [27, 28]  0.000133  0.004339      1  1437
12   0.0716  0.3124        []  0.000133  0.002979      1   301
21   0.1278  0.9812        []  0.000425  0.045276      2   837
22   0.1278  0.3882  [29, 30]  0.000425  0.004584      2  1410
27   0.3768  0.4244  [31, 32]  0.004339  0.005586     11   863
28   0.3768  1.0000        []  0.004339  0.071075     11   406
29   0.3882  0.9292        []  0.004584  0.032849     22   262
30   0.3882  0.9786        []  0.004584  0.043969     22   668
31   0.4244  0.9896        []  0.005586  0.048706     27   428
32   0.4244  0.9992        []  0.005586  0.064437     27   395

\end{verbatim}

\end{codeoutput}
\end{codecell}
The next step is to assign cluster labels to a set of foreground data
points with the function \texttt{GeomTree.getClusterLabels}. The desired
labeling method is specified with the \texttt{method} argument. When the
correct number of clusters $K$ is known, the \texttt{first-k} option
retrieves the first $K$ disjoint clusters that appear when $\lambda$ is
increased from 0. Alternately, the \texttt{upper-set} option cuts the
tree at a single level, which is useful if the goal is to include or
exclude a certain fraction of the data from the upper level set. Here we
use this function with $\alpha$ set to 0.05, which removes the 5\% of
the observations with the lowest estimated density (i.e.~outliers) and
clusters the remainder. Finally, the \texttt{all-mode} option returns a
foreground cluster for each leaf of the level set tree, which avoids the
need to specify either $K$, $\lambda$, or $\alpha$.

Additional arguments for each method are specified by keyword argument;
the \texttt{getClusterLabels} method parses them intelligently. For all
of the labeling methods the function returns two objects. The first is
an $m \times 2$ matrix, where $m$ is the number of points in the
foreground set. The first column is the index of an observation in the
full data matrix, and the second column is the cluster label. The second
object is a list of the tree nodes that are foreground clusters. This is
useful for coloring level set tree nodes to match observations plotted
in feature space.

\begin{figure}
    \centering
    \begin{subfigure}[b]{0.45\textwidth}
        \centering
        \includegraphics[width=\textwidth]{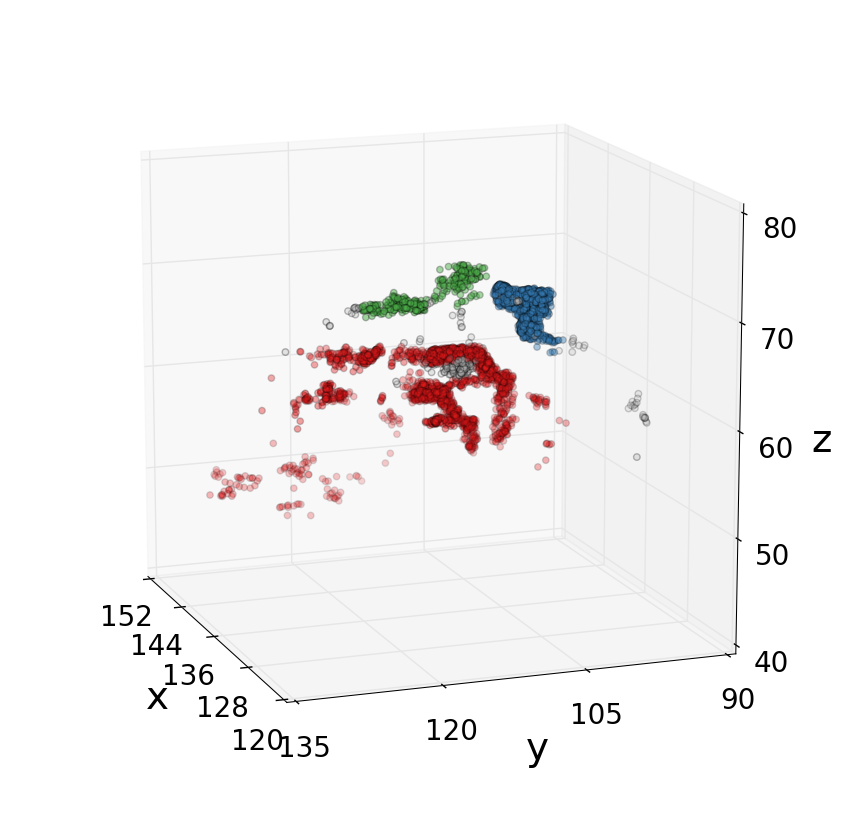}
        \caption{Fiber endpoint data, colored by \textit{first-K} foreground cluster}
        \label{fig:endpt_firstK_fg}
    \end{subfigure}
    \begin{subfigure}[b]{0.45\textwidth}
        \centering
        \includegraphics[width=\textwidth]{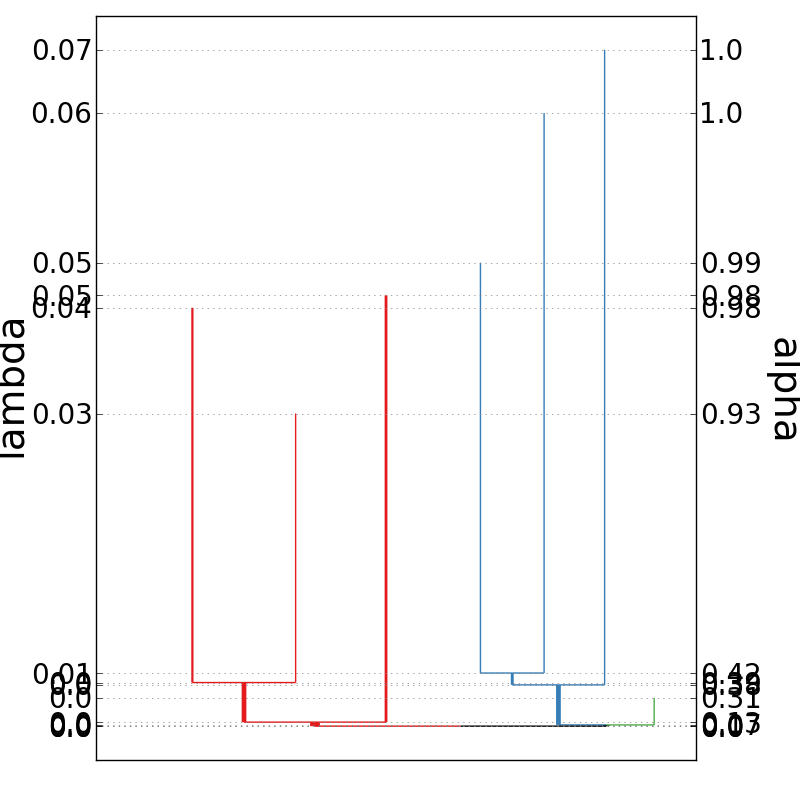}
        \caption{Lambda scale}
        \label{fig:lambda_tree}
    \end{subfigure}
    \begin{subfigure}[b]{0.45\textwidth}
        \centering
        \includegraphics[width=\textwidth]{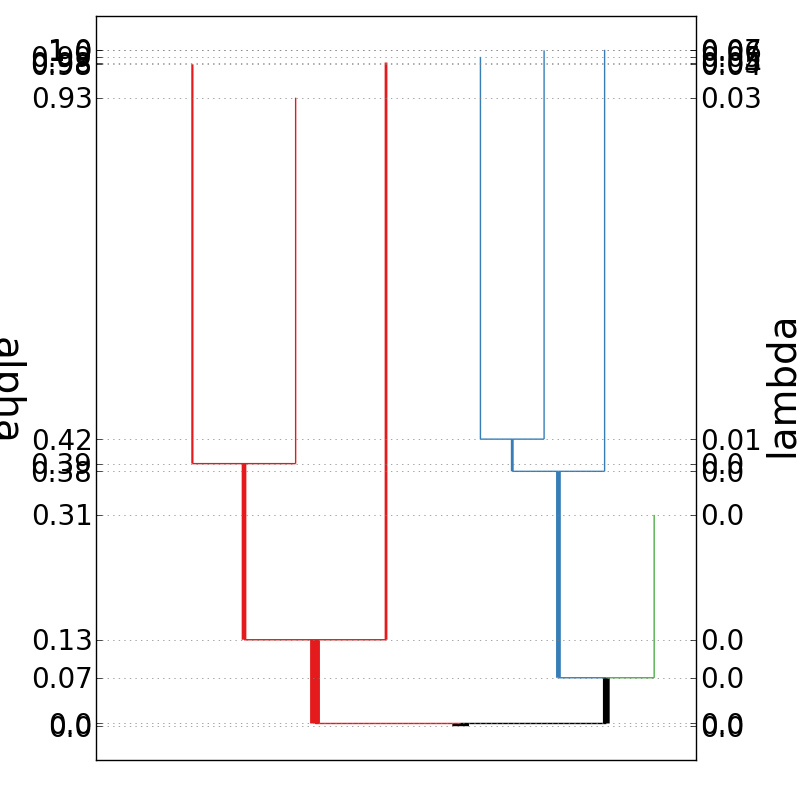}
        \caption{Alpha scale}
        \label{fig:alpha_tree}
    \end{subfigure}
    \begin{subfigure}[b]{0.45\textwidth}
        \centering
        \includegraphics[width=\textwidth]{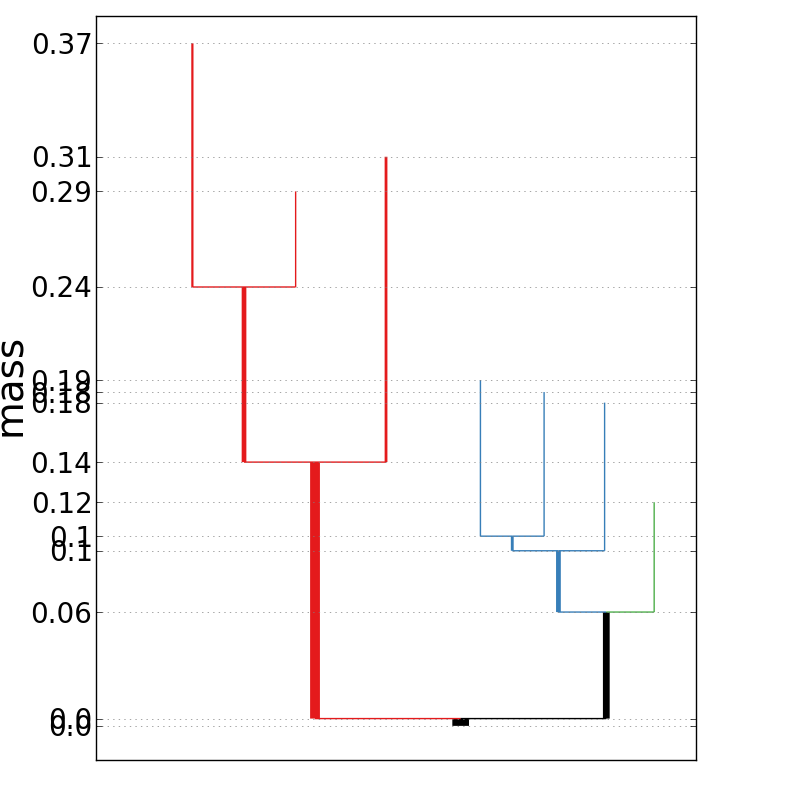}
        \caption{Kappa scale}
        \label{fig:kappa_tree}
    \end{subfigure}
    \caption{First-k clustering results with different vertical scales and the clusters in feature space.}
    \label{fig:firstK_trees}
\end{figure}

\begin{codecell}
\begin{codeinput}
\begin{lstlisting}
uc_k, nodes_k = tree.getClusterLabels(method='first-k', k=3)
uc_lambda, nodes_lambda = tree.getClusterLabels(method='upper-set', threshold=0.05, scale='lambda')
uc_mode, nodes_mode = tree.getClusterLabels(method='all-mode')
\end{lstlisting}
\end{codeinput}

\end{codecell}
The \texttt{GeomTree.plot} method draws the level set tree dendrogram,
with the vertical scale controlled by the \texttt{form} parameter. See
Section \ref{ssec:scales} for more detail. The three plot forms are
shown in Figure \ref{fig:firstK_trees}, foreground clusters are derived
from \textit{first-k} clustering with $K$ set to 3. the
\texttt{plotForeground} function from the \pkg{DeBaCl} \pkg{utils}
module is used to match the node colors in the dendrogram to the
clusters in feature space. Note that the plot function returns a tuple
with several objects, but only the first is useful for most
applications.

\begin{codecell}
\begin{codeinput}
\begin{lstlisting}
## Plot the level set tree with three different vertical scales, colored by the first-K clustering
fig = tree.plot(form='lambda', width='mass', color_nodes=nodes_k)[0]
fig.savefig('../figures/endpt_tree_lambda.png')

fig = tree.plot(form='alpha', width='mass', color_nodes=nodes_k)[0]
fig.savefig('../figures/endpt_tree_alpha.png')

fig = tree.plot(form='kappa', width='mass', color_nodes=nodes_k)[0]
fig.savefig('../figures/endpt_tree_kappa.png')


## Plot the foreground points from the first-K labeling
fig, ax = utl.plotForeground(X, uc_k, fg_alpha=0.6, bg_alpha=0.4, edge_alpha=0.3, s=22)
ax.elev = 14; ax.azim=160    # adjust the camera angle
fig.savefig('../figures/endpt_firstK_fg.png', bbox_inches='tight')
\end{lstlisting}
\end{codeinput}

\end{codecell}

\begin{figure}
    \centering
    \begin{subfigure}[b]{0.45\textwidth}
        \centering
        \includegraphics[width=\textwidth]{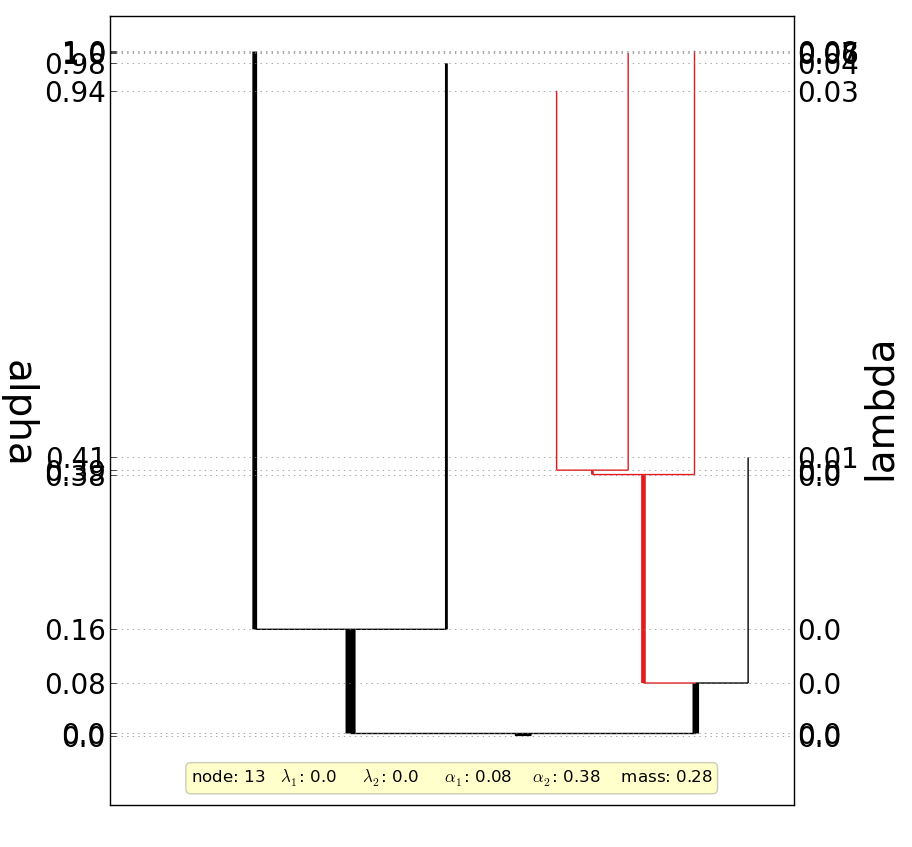}
        \caption{}
        \label{fig:endpt_subtree}
    \end{subfigure}
    \begin{subfigure}[b]{0.45\textwidth}
        \centering
        \includegraphics[width=\textwidth]{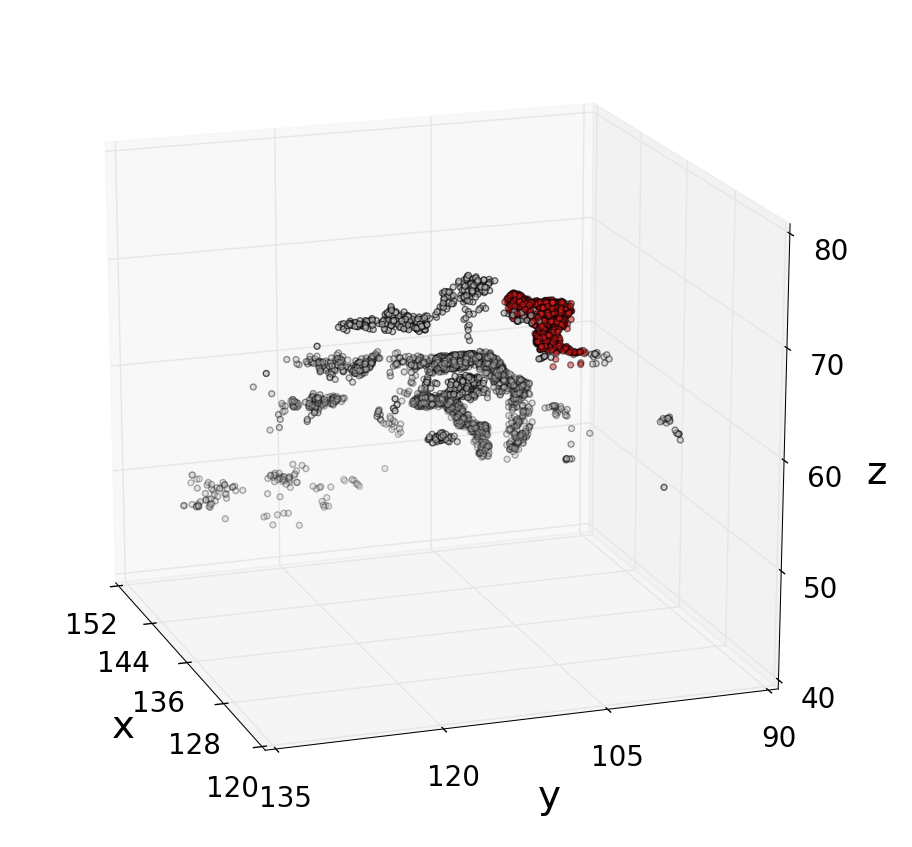}
        \caption{}
        \label{fig:endpt_subset}
    \end{subfigure}
    \begin{subfigure}[b]{0.45\textwidth}
        \centering
        \includegraphics[width=\textwidth]{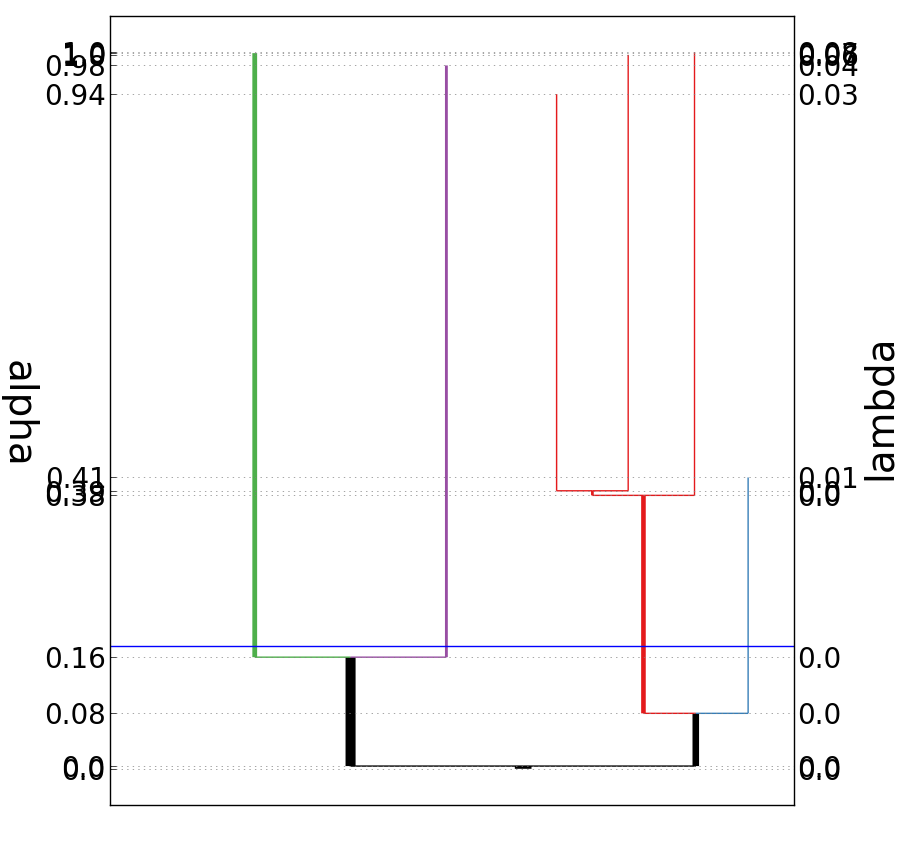}
        \caption{}
        \label{fig:endpt_alpha20_tree}
    \end{subfigure}
    \begin{subfigure}[b]{0.45\textwidth}
        \centering
        \includegraphics[width=\textwidth]{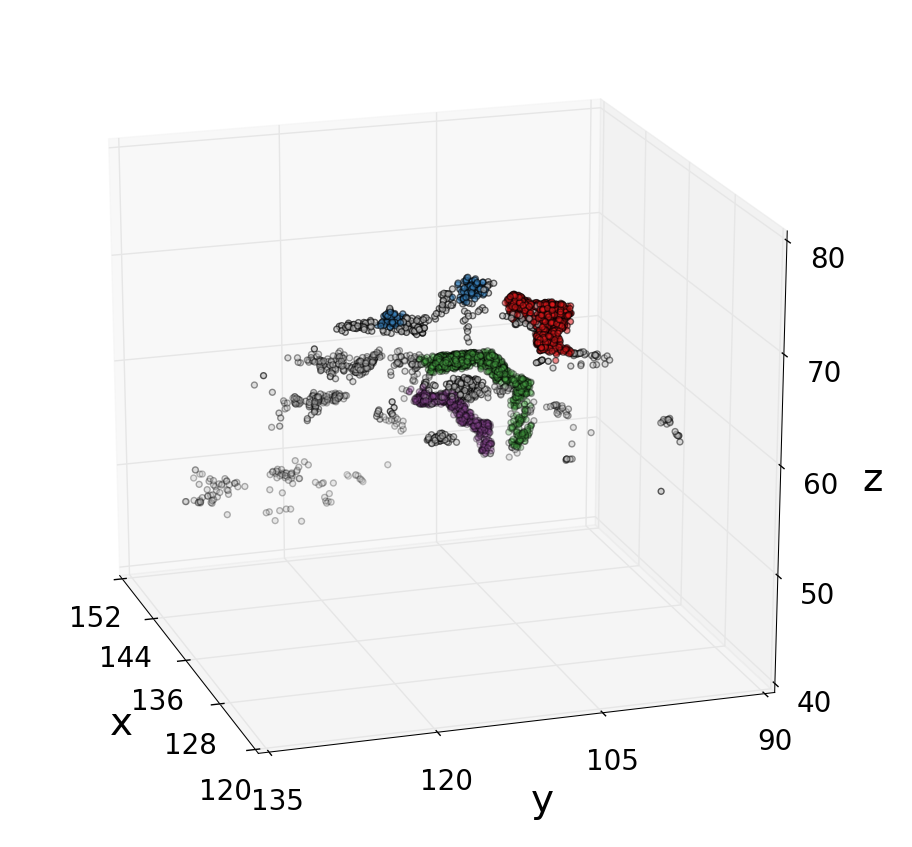}
        \caption{}
        \label{fig:endpt_alpha20}
    \end{subfigure}
    \caption{The level set tree can be used as a scaffold for interactive exploration of data subsets or upper level set clusters.}
    \label{fig:gui_tools}
\end{figure}

A level set tree plot is also useful as a scaffold for interactive
exploration of spatially coherent subsets of data, either by selecting
individual nodes of the tree or by retreiving high-density clusters at a
selected density or mass level. These tools are particularly useful for
exploring clustering features at multiple data resolutions. In Figure
\ref{fig:gui_tools}, for example, there are two dominant clusters, but
each one has highly salient clustering behavior at higher resolutions.
The interactive tools allow for exploration of the parent-child
relationships between these clusters.

\begin{codecell}
\begin{codeinput}
\begin{lstlisting}
tool1 = gtree.ComponentGUI(tree, X, form='alpha', output=['scatter'], size=18, width='mass')
tool1.show()

tool2 = gtree.ClusterGUI(tree, X, form='alpha', width='mass', size=18)
tool2.show()
\end{lstlisting}
\end{codeinput}

\end{codecell}

\begin{figure}
    \centering
    \includegraphics[width=0.65\textwidth]{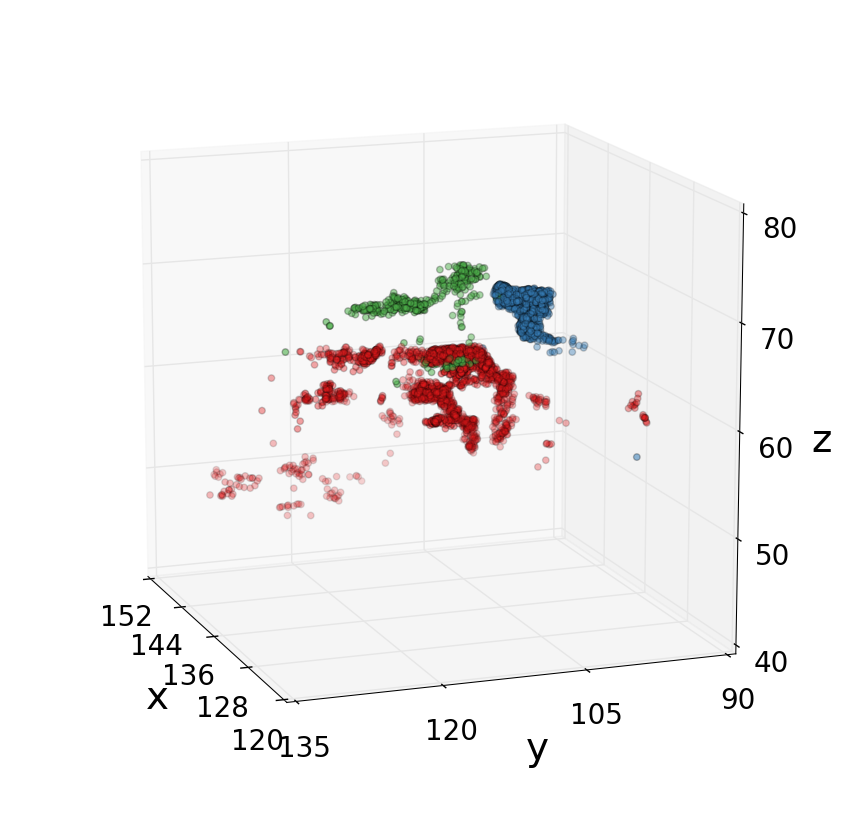}
    \caption{Endpoint data, with background points assigned to the \textit{first-K} foreground clusters with a k-nearest neighbor classifier.}
    \label{fig:endpt_segment}
\end{figure}

The final step of our standard data analysis is to assign background
points to a foreground cluster. \pkg{DeBaCl}'s \pkg{utils} module
includes several very simple classifiers for this task, although more
sophisticated methods have been proposed \citep{Azzalini2007}. For this
example we assign background points with a k-nearest neighbor
classifier. The observations are plotted a final time, with a full data
partition (Figure \ref{fig:endpt_segment}).

\begin{codecell}
\begin{codeinput}
\begin{lstlisting}
## Assign background points with a simple kNN classifier
segment = utl.assignBackgroundPoints(X, uc_k, method='knn', k=k)

## Plot all observations, colored by cluster
fig, ax = utl.plotForeground(X, segment, fg_alpha=0.6, bg_alpha=0.4, edge_alpha=0.3, s=22)
ax.elev = 14; ax.azim=160
fig.savefig('../figures/endpt_firstK_segment.png', bbox_inches='tight')
\end{lstlisting}
\end{codeinput}

\end{codecell}

To customize the level set tree estimator, each phase can be done
manually. Here we use methods in \pkg{DeBaCl}'s \pkg{utils} module to
build a k-nearest neighbor similarity graph \texttt{W}, a k-nearest
neighbor density estimate \texttt{fhat}, a grid of density levels
\texttt{levels}, and the background observation sets at each density
level (\texttt{bg\_sets}). The \texttt{constructTree} method of the
\pkg{geom\_tree} module puts the pieces together to make the tree and
the \texttt{prune} function removes tree leaf nodes that are small and
likely due to random noise.

\begin{codecell}
\begin{codeinput}
\begin{lstlisting}
## Similarity graph and density estimate
W, k_radius = utl.knnGraph(X, k, self_edge=False)
fhat = utl.knnDensity(k_radius, n, p, k)

## Tree construction and pruning
bg_sets, levels = utl.constructDensityGrid(fhat, mode='mass', n_grid=None)
tree = gtree.constructTree(W, levels, bg_sets, mode='density', verbose=False)
tree.prune(method='size-merge', gamma=gamma)
print tree

\end{lstlisting}
\end{codeinput}
\begin{codeoutput}

\begin{verbatim}
     alpha1  alpha2  children   lambda1   lambda2 parent  size
key                                                           
0    0.0000  0.0040    [1, 2]  0.000000  0.000003   None  5000
1    0.0040  0.0716  [11, 12]  0.000003  0.000133      0  2030
2    0.0040  0.1278  [21, 22]  0.000003  0.000425      0  2950
11   0.0716  0.3768  [27, 28]  0.000133  0.004339      1  1437
12   0.0716  0.3124        []  0.000133  0.002979      1   301
21   0.1278  0.9812        []  0.000425  0.045276      2   837
22   0.1278  0.3882  [29, 30]  0.000425  0.004584      2  1410
27   0.3768  0.4244  [31, 32]  0.004339  0.005586     11   863
28   0.3768  1.0000        []  0.004339  0.071075     11   406
29   0.3882  0.9292        []  0.004584  0.032849     22   262
30   0.3882  0.9786        []  0.004584  0.043969     22   668
31   0.4244  0.9896        []  0.005586  0.048706     27   428
32   0.4244  0.9992        []  0.005586  0.064437     27   395

\end{verbatim}

\end{codeoutput}
\end{codecell}
In the definition of density levels and background sets, the
\texttt{constructDensityGrid} allows the user to specify the
\texttt{n\_grid} parameter to speed up the algorithm by computing the
upper level set and connectivity for only a subset of density levels.
The \texttt{mode} parameter determines whether the grid of density
levels is based on evenly-sized blocks of observations
(\texttt{mode='mass'}) or density levels (\texttt{mode='levels'}); we
generally prefer the `mass' mode for our own analyses.

The \texttt{mode} parameter of the tree construction function is usually
set to be `density', which treats the underlying function \texttt{fhat}
as a density or pseudo-density function, with a floor value of 0. This
algorithm can be applied to arbitrary functions that do not have a floor
value, in which case the mode should be set to `general'.


\subsection{Extension: The Chaudhuri-Dasgupta Tree}\label{ssec:extend}

Usage of the Chaudhuri-Dasgupta algorithm is similar to the standalone
\texttt{geomTree} function. First load the \pkg{DeBaCl} module
\pkg{cd\_tree} (labeled here for brevity as \pkg{cdt}) and the utility
functions in \pkg{utils}, as well as the data.

\begin{codecell}
\begin{codeinput}
\begin{lstlisting}
## Import DeBaCl package
import sys
sys.path.append('/home/brian/Projects/debacl/DeBaCl/')

from debacl import cd_tree as cdt
from debacl import utils as utl

## Import other Python libraries
import numpy as np

## Load the data
X = np.loadtxt('0187_endpoints.csv', delimiter=',')
n, p = X.shape
\end{lstlisting}
\end{codeinput}

\end{codecell}
Because the straightforward implementation of the Chaudhuri-Dasgupta
algorithm is extremely slow, we use a random subset of only 200
observations (out of the total of 10,000). The smoothing parameter is
set to be 2.5\% of $n$, or 5. The pruning parameter is 5\% of $n$, or
10. The pruning parameter is slightly less important for the
Chaudhuri-Dasgupta algorithm.

\begin{codecell}
\begin{codeinput}
\begin{lstlisting}
## Downsample
n_samp = 200
ix = np.random.choice(range(n), size=n_samp, replace=False)
X = X[ix, :]
n, p = X.shape

## Set level set tree parameters
p_k = 0.025
p_gamma = 0.05
k = int(p_k * n)
gamma = int(p_gamma * n)

## Set plotting parameters
utl.setPlotParams(axes_labelsize=28, xtick_labelsize=20, ytick_labelsize=20, figsize=(8,8))
\end{lstlisting}
\end{codeinput}

\end{codecell}
The straightforward implementation of the Chaudhuri-Dasgupta algorithm
starts with a complete graph and removes one edge a time, which is
extremely slow. The \texttt{start} parameter of the \texttt{cdTree}
function allows for shortcuts. These are approximations to the method,
but are necessary to make the algorithm practical. Currently, the only
implemented shortcut is to start with a k-nearest neighbor graph.

\begin{codecell}
\begin{codeinput}
\begin{lstlisting}
## Construct the level set tree estimate
tree = cdt.cdTree(X, k, alpha=1.4, start='knn', verbose=False)
tree.prune(method='size-merge', gamma=gamma)
\end{lstlisting}
\end{codeinput}

\end{codecell}
As with the geometric tree, we can print a summary of the tree, plot the
tree, retrieve foreground cluster labels, and plot the foreground
clusters. This is illustrated below for the `all-mode' labeling method.

\begin{codecell}
\begin{codeinput}
\begin{lstlisting}
## Print/make output
print tree

fig = tree.plot()
fig.savefig('../figures/cd_tree.png')

uc, nodes = tree.getClusterLabels(method='all-mode')

fig, ax = utl.plotForeground(X, uc, fg_alpha=0.6, bg_alpha=0.4, edge_alpha=0.3, s=60)
ax.elev = 14; ax.azim=160
fig.savefig('../figures/cd_allmode.png')
\end{lstlisting}
\end{codeinput}
\begin{codeoutput}

\begin{verbatim}
     children parent        r1        r2  size
key                                           
0      [3, 4]   None  8.134347  4.374358   200
3    [15, 16]      0  4.374358  2.220897   109
4    [23, 24]      0  4.374358  1.104121    75
15         []      3  2.220897  0.441661    32
16         []      3  2.220897  0.343408    55
23         []      4  1.104121  0.445529    28
24         []      4  1.104121  0.729226    24

\end{verbatim}

\end{codeoutput}
\end{codecell}
\begin{figure}
    \centering
    \begin{subfigure}[b]{0.49\textwidth}
        \centering
        \includegraphics[width=0.75\textwidth]{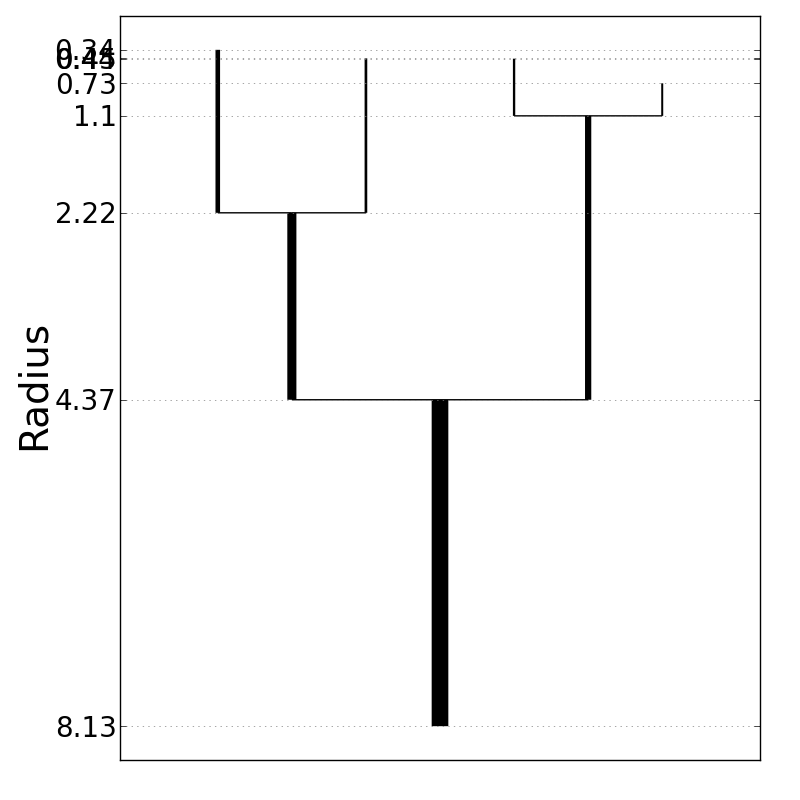}
        \caption{}
        \label{fig:cd_tree}
    \end{subfigure}
    \begin{subfigure}[b]{0.49\textwidth}
        \centering
        \includegraphics[width=0.75\textwidth]{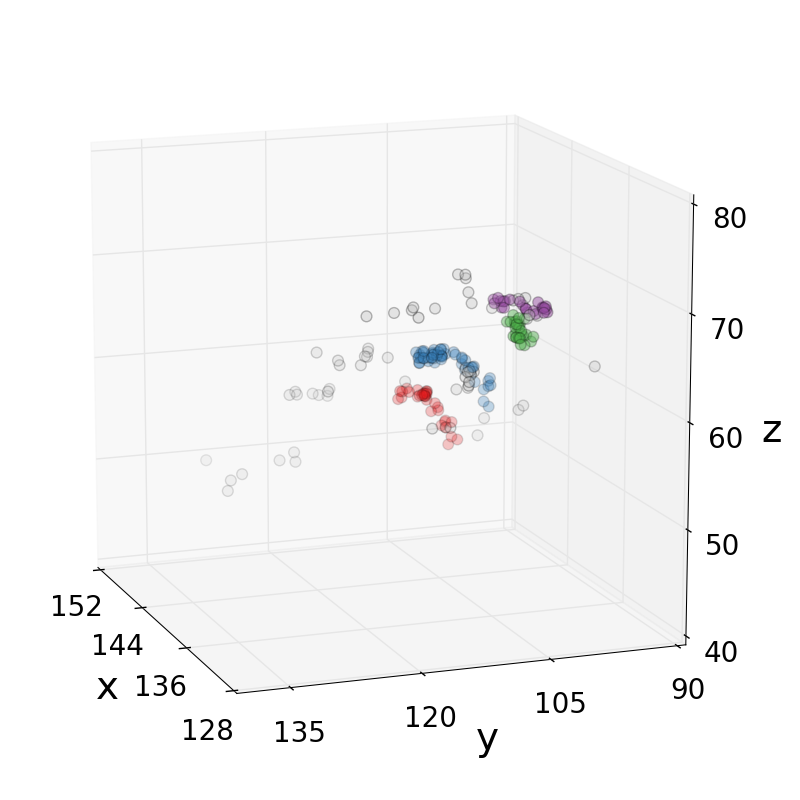}
        \caption{}
        \label{fig:cd_allmode}
    \end{subfigure}
    \caption{The Chaudhuri-Dasgupta tree for the fiber track endpoint data, downsampled from 10,000 to 200 observations to make computation feasible. Foreground clusters based on all-mode clustering are shown on the right.}
    \label{fig:dasgupta}
\end{figure}


\subsection{Extension: Functional Data}\label{ssec:fda}

Nothing in the process of estimating a level set tree requires
$\widehat{f}$ to be a bona fide probability density function, and the
\pkg{DeBaCl} package allows us to use this fact to use level set trees
for much more complicated datasets. To illustrate we use the phoneme
dataset from Ferraty and Vieu \citeyearpar{Ferraty2006}, which contains
2000 total observations of five short speech patterns. Each observation
is recorded on a regular grid of 150 frequencies, but we treat this as
an approximation of a continuous function on an interval of
$\mathbb{R}^1$. Because the observations are random curves they do not
have bona fide density functions, but we can still construct a sample
level set tree by estimating a pseudo-density function that measures the
proximity of each curve to its neighbors.

\begin{figure}
    \centering
    \includegraphics[width=0.65\textwidth]{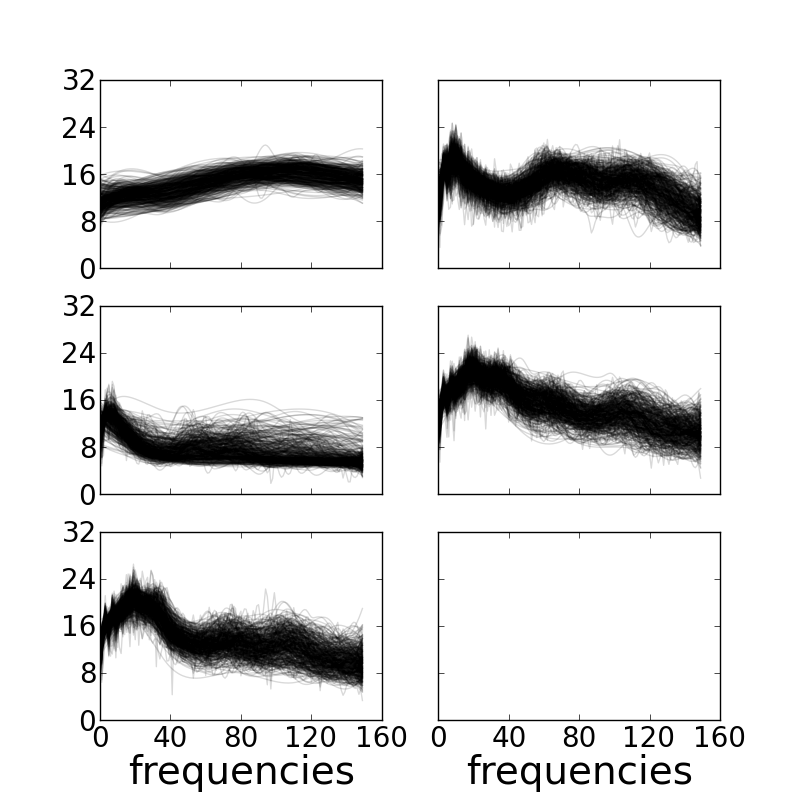}
    \caption{Smoothed waveforms for spoken phonemes, separated by true phoneme.}
    \label{fig:phoneme_data}
\end{figure}

To start we load the \pkg{DeBaCl} modules and the data, which have been
pre-smoothed for this example with cubic splines. The true class of each
observation is in the last column of the raw data object. The curves for
each phoneme are shown in Figure \ref{fig:phoneme_data}.

\begin{codecell}
\begin{codeinput}
\begin{lstlisting}
## Import DeBaCl package
import sys
sys.path.append('/home/brian/Projects/debacl/DeBaCl/')

from debacl import geom_tree as gtree
from debacl import utils as utl

## Import other Python libraries
import numpy as np
import scipy.spatial.distance as spdist
import matplotlib.pyplot as plt

## Set plotting parameters
utl.setPlotParams(axes_labelsize=28, xtick_labelsize=20, ytick_labelsize=20, figsize=(8,8))

## Load data
speech = np.loadtxt('smooth_phoneme.csv', delimiter=',')
phoneme = speech[:, -1].astype(np.int)
speech = speech[:, :-1]

n, p = speech.shape
\end{lstlisting}
\end{codeinput}

\end{codecell}
\begin{codecell}
\begin{codeinput}
\begin{lstlisting}
## Plot the curves, separated by true phoneme
fig, ax = plt.subplots(3, 2, sharex=True, sharey=True)
ax = ax.flatten()
ax[-2].set_xlabel('frequencies')
ax[-1].set_xlabel('frequencies')

for g in np.unique(phoneme):
	ix = np.where(phoneme == g)[0]
	for j in ix:
		ax[g].plot(speech[j, :], c='black', alpha=0.15)
        
fig.savefig('../figures/phoneme_data.png')
\end{lstlisting}
\end{codeinput}

\end{codecell}

For functional data we need to define a distance function, precluding
the use of the convenience method \texttt{GeomTree.geomTree} or even the
utility function \texttt{utils.knnGraph}. First the bandwith and tree
pruning parameters are set to be $0.01n$. In the second step all
pairwise distances are computed in order to find the $k$-nearest
neighbors for each observation. For simplicity we use Euclidean distance
between a pair of curves (which happens to work well in this example),
but this is not generally optimal. Next, the adjacency matrix for a
$k$-nearest neighbor graph is constructed, which is no different than
the finite-dimensional case. Finally the pseudo-density estimator is
built by using the finite-dimenisonal $k$-nearest neighbor density
estimator with the dimension set (incorrectly) to 1. This function does
not integrate to 1, but the function induces an ordering on the
observations (from smallest to largest $k$-neighbor radius) that is
invariant to the dimension. This ordering is all that is needed for the
final step of building the level set tree.

\begin{codecell}
\begin{codeinput}
\begin{lstlisting}
## Bandwidth and pruning parameters
p_k = 0.01
p_gamma = 0.01
k = int(p_k * n)
gamma = int(p_gamma * n)

## Find all pairwise distances and the indices of each point's k-nearest neighbors
D = spdist.squareform(spdist.pdist(speech, metric='euclidean'))
rank = np.argsort(D, axis=1)
ix_nbr = rank[:, 0:k]
ix_row = np.tile(np.arange(n), (k, 1)).T

## Construct the similarity graph adjacency matrix
W = np.zeros(D.shape, dtype=np.bool)
W[ix_row, ix_nbr] = True
W = np.logical_or(W, W.T)
np.fill_diagonal(W, False)

## Compute a pseudo-density estimate and evaluate at each observation
k_nbr = ix_nbr[:, -1]
r_k = D[np.arange(n), k_nbr]
fhat = utl.knnDensity(r_k, n, p=1, k=k)

## Build the level set tree
bg_sets, levels = utl.constructDensityGrid(fhat, mode='mass', n_grid=None)
tree = gtree.constructTree(W, levels, bg_sets, mode='density', verbose=False)
tree.prune(method='size-merge', gamma=gamma)
print tree
\end{lstlisting}
\end{codeinput}
\begin{codeoutput}

\begin{verbatim}
     alpha1  alpha2 children   lambda1   lambda2 parent  size
key                                                          
0    0.0000  0.2660   [1, 2]  0.000000  0.000261   None  2000
1    0.2660  0.3435   [3, 4]  0.000261  0.000275      0  1125
2    0.2660  1.0000       []  0.000261  0.000938      0   343
3    0.3435  0.4905   [5, 6]  0.000275  0.000307      1   565
4    0.3435  0.7705       []  0.000275  0.000426      1   413
5    0.4905  0.9920       []  0.000307  0.000808      3   391
6    0.4905  0.7110       []  0.000307  0.000382      3    85

\end{verbatim}

\end{codeoutput}
\end{codecell}
Once the level set tree is constructed we can plot it and retrieve
cluster labels as with finite-dimensional data. In this case we choose
the all-mode cluster labeling which produces four clusters. The utility
function \texttt{utils.plotForeground} is currently designed to work
only with two- or three-dimensional data, so plotting the foreground
clusters must be done manually for functional data. The clusters from
this procedure match the true groups quite well, at least in a
qualitative sense.

\begin{codecell}
\begin{codeinput}
\begin{lstlisting}
## Retrieve cluster labels
uc, nodes = tree.getClusterLabels(method='all-mode')

## Level set tree plot
fig = tree.plot(form='alpha', width='mass', color_nodes=nodes)[0]
fig.savefig('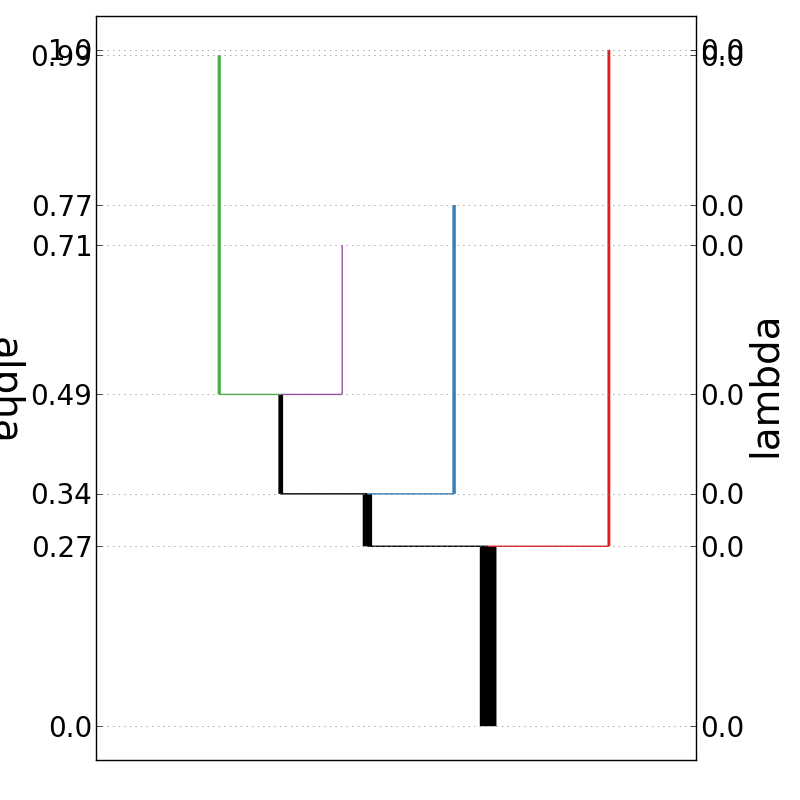')

## Plot the curves, colored by foreground cluster
palette = utl.Palette()
fig, ax = plt.subplots()
ax.set_xlabel("frequency index")

for c in np.unique(uc[:,1]):
    ix = np.where(uc[:,1] == c)[0]
    ix_clust = uc[ix, 0]
	
    for i in ix_clust:
        ax.plot(speech[i,:], c=np.append(palette.colorset[c], 0.25))

fig.savefig('../figures/phoneme_allMode.png')
\end{lstlisting}
\end{codeinput}

\end{codecell}
\begin{figure}
    \centering
    \includegraphics[width=0.65\textwidth]{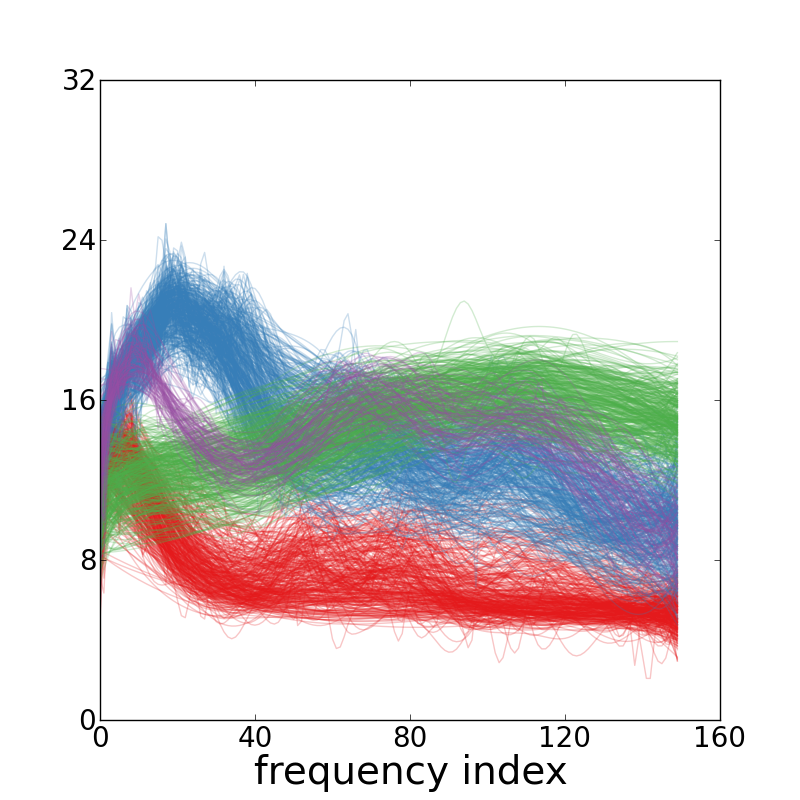}
    \caption{All-mode foreground clusters for the smoothed phoneme data.}
    \label{fig:phoneme_allMode}
\end{figure}


\section{Conclusion}\label{sec:conc}
The Python package \pkg{DeBaCl} for hierarchical density-based clustering provides a highly usable implementation of level set tree estimation and clustering. It improves on existing software through computational efficiency and a high-degree of modularity and customization. Namely, \pkg{DeBaCl}:
\begin{itemize}
	\item offers the first known implementation of the theoretically
	well-supported Chaudhuri-Dasgupta level set tree algorithm;
	
	\item allows for very general data ordering functions, which are typically
	probability density estimates but could also be pseudo-density estimates for
	infinite-dimensional functional data or even arbitrary functions;
	
	\item accepts any similarity graph, density estimator, pruning function,
	cluster labeling scheme, and background point assignment classifier;
	
	\item includes the all-mode cluster labeling scheme, which does not require
	an \textit{a priori} choice of the number of clusters;
	
	\item incorporates the $\lambda$, $\alpha$, and $\kappa$ vertical scales for
	plotting level set trees, as well as other plotting tweaks to make level set
	tree plots more interpretable and usable;
	
	\item and finally, includes interactive GUI tools for selecting coherent
	data subsets or high-density clusters based on the level set tree.
\end{itemize}

The \pkg{DeBaCl} package and user manual is available at
\url{https://github.com/CoAxLab/DeBaCl}. The project remains under active
development; the focus for the next version will be on improvements in
computational efficiency, particularly for the Chaudhuri-Dasgupta algorithm.

\section*{Acknowledgments}
This research was sponsored by the Army Research Laboratory and was accomplished
under Cooperative Agreement Number W911NF-10-2-0022. The views and conclusions
contained in this document are those of the authors and should not be
interpreted as representing the official policies, either expressed or implies,
of the Army Research Laboratory or the U.S. Government. The U.S. Government is
authorized to reproduce and distribute reprints for the Government purposes
notwithstanding any copyright notation herein. This research was also supported
by NSF CAREER grant DMS 114967.

\bibliography{biblio/densclust,biblio/functional,biblio/books,biblio/genclust}

\end{document}